\documentclass[10pt, conference]{IEEEtran}

\usepackage[utf8]{inputenc} 
\usepackage[T1]{fontenc}
\usepackage[cmex10]{amsmath} 
\usepackage{epsfig,rotating,setspace,latexsym,amsmath,epsf,amssymb,amsfonts,bm,theorem,cite,enumerate,longtable,accents,float,physics}
\usepackage{algorithm,algorithmic,graphicx,epsf,authblk,epstopdf,url,xcolor, soul,multirow,bbm}
\usepackage{mathtools,comment}
\usepackage[center]{qtree}
\usepackage{tree-dvips}
\usepackage[linguistics]{forest }

\newtheorem{theorem}{Theorem}
\newtheorem{corollary}{Corollary}

\newtheorem{remark}{Remark}
\newtheorem{lemma}{Lemma}

\newenvironment{Proof}[1]{\medskip\par\noindent{\bf Proof:\,}\,#1}{{\mbox{\,$\blacksquare$}\par}}

\newcommand{\cq}{{\mathcal{Q}}}

\newcommand{\cB}{{\mathcal{B}}}

\title{The Asymptotic Capacity of Byzantine Symmetric Private Information Retrieval and Its Consequences}
\author{Mohamed Nomeir \qquad Alptug Aytekin \qquad Sennur Ulukus\\
	\normalsize Department of Electrical and Computer Engineering\\
	\normalsize University of Maryland, College Park, MD 20742 \\
	\normalsize \emph{mnomeir@umd.edu} \qquad      \emph{aaytekin@umd.edu} \qquad
         \emph{ulukus@umd.edu}}
 
\begin{document}

\maketitle

\begin{abstract}
  We consider the problem of finding the asymptotic capacity of symmetric private information retrieval (SPIR) with $B$ Byzantine servers. Prior to finding the capacity, a definition for the Byzantine servers is needed since in the literature there are two different definitions. In \cite{byzantine_tpir}, where it was first defined, the Byzantine servers can send any symbol from the storage, their received queries and some independent random symbols. In \cite{unresponsive_byzantine_1}, Byzantine servers send any random symbol independently of their storage and queries. It is clear that these definitions are not identical, especially when \emph{symmetric} privacy is required. To that end, we define Byzantine servers, inspired by \cite{byzantine_tpir}, as the servers that can share everything, before and after the scheme initiation. In this setting, we find an upper bound, for an infinite number of messages case, that should be satisfied for all schemes that protect against this setting and develop a scheme that achieves this upper bound. Hence, we identify the capacity of the problem.
\end{abstract}

\section{Introduction}
The problem of private information retrieval (PIR) was first introduced in \cite{chor}. In PIR, a user wishes to retrieve a message out of $K$ messages from $N$ replicated non-colluding servers. It was shown in \cite{c_pir} that the capacity, i.e., the highest possible ratio between the required message length and the total downloaded symbols is $C_{PIR}(N,K) = (1+\frac{1}{N}+ \ldots+\frac{1}{N^{K-1}})^{-1}$. A variation was developed afterward where the user cannot get any information about the other messages except the required message index, coined as \emph{symmetric} PIR (SPIR). It was shown in \cite{c_spir} that the capacity is $C_{SPIR}(N,K) = \lim_{K\rightarrow \infty}C_{PIR}(N,K)= 1 - \frac{1}{N}$. The $T$-colluding PIR (TPIR) was first defined in \cite{c_tpir}, where any $T < N$ servers can collude, i.e., share the queries transmitted by the user. It was shown that the capacity is equal to $C_{TPIR}(T,N,K) = C_{PIR}(\frac{N}{T},K)$. In addition, the symmetric TPIR (TSPIR) capacity is given by $C_{TSPIR}(T,N,K) = C_{SPIR}(\frac{N}{T},K)$ as shown in \cite{tspir_mdscoded}. If instead of uniform collusion, the user has knowledge about the collusion pattern among the servers, the capacity for this setting is found in \cite{arbitrarycollusion}. In \cite{sun_eaves}, the model for passive eavesdropper TPIR was developed, where there are $E$ links contaminated with passive eavesdroppers who can listen to both the queries and the answers. It was shown that the capacity for this setting is $C_{ETPIR}(E,T,N,K) = C_{PIR}(\frac{N}{M},K)$, where $M = \max\{T,E\} < N$. The symmetric capacity for the same setting was found in \cite{c_TESPIR}, which is $C_{ETSPIR}(E,T,N,K) = C_{SPIR}(\frac{N}{M},K)$. For arbitrary eavesdropper patterns, the capacity was found in \cite{nan_eaves}. The $X$ secure TPIR was first defined in \cite{first_xsecure}, where any $X$ servers are allowed to communicate their storage. The asymptotic capacity was found in \cite{csa} as $\lim_{K \rightarrow \infty}C_{XTPIR}(X,T,N,K) = 1- \frac{X+T}{N}$. 

Another variation, of utmost importance for our paper, is the Byzantine TPIR (BTPIR). As mentioned previously, it was first defined in \cite{byzantine_tpir}, where any $B$ servers can send any symbol from their storage, query, or any arbitrary random variables of their choice. It was shown that the capacity of this setting is $C_{BTPIR}(B,T,N,K) = \frac{N-2B}{N} C_{TPIR}(T,N-2B,K)=\frac{N-2B}{N} C_{PIR}(\frac{N-2B}{T},K)$. In \cite{unresponsive_byzantine_1}, another definition of the Byzantine servers was defined in the setting of the symmetric BTPIR (BTSPIR), where instead of the ability of the Byzantine servers to send any arbitrary answer based on their storage and queries, and potentially the masking random variable, it was assumed that the Byzantine servers can send only a symbol independent of the storage and the queries. Although the approach is understandable since it mitigates errors in the channel as a special case for this setting, it restricts the capabilities of the Byzantine servers as defined in \cite{byzantine_tpir} since the model does not allow the Byzantine servers to send anything related to their storage or received queries. 

In this paper, we define a new model for the Byzantine servers, where they are allowed to share the storage, masking random variables and queries. In addition, they can agree on a plan to affect the system's integrity. In this way, the definition is more general and follows the definition in \cite{byzantine_tpir}. We develop a scheme that can combat this Byzantine collaborative behavior and provide symmetric security as well. In addition, we provide an upper bound on the rate for this setting for infinite number of messages, and show that the developed scheme achieves this upper bound. This result is captured in the main theorem in this paper; however, the consequences of this result are as significant as the result itself. For example, the method developed to provide the masking scheme provides better rates for quantum Byzantine with static or dynamic eavesdroppers $T$ colluding $X$ secure symmetric PIR found in \cite{our_symmetric_quantum_byzantine_Eaves_journal}. For more work related to PIR, see \cite{AG_1, AG_2, banawan_eaves, banawan_multimessage_pir, banawan_pir_mdscoded, batuhan_hetero, ChaoTian, codedstorage_adversary_tpir, wei_banawan_cache_pir, ulukusPIRLC, wang_spir, nomeirasymmetric}.

\section{Problem Formulation}
Let $N$ be the number of servers and $W_{[K]}$ be the set of messages available in the system, each of length $L$. There exists $\mathcal{B} \subset [N]$, such that $|\mathcal{B}| \leq B$ are Byzantine servers. The Byzantine servers \emph{share an open communication channel}, i.e., they can communicate anything to each other. The party holding the messages, $W_{[K]}$, wants to distribute these messages such that any action of the Byzantine servers will not leak information about the messages, therefore, none of these Byzantine servers should know the messages to ensure symmetric privacy, i.e., if Byzantine servers send $A_1, \ldots, A_B$, none of the messages is leaked, i.e., 
\begin{align}\label{pf_sec_1}
    I(A_{[B]};W_{[K]}) = 0. 
\end{align} 

This condition clearly implies that the storage should be secure against any $B$ communications, i.e., 
\begin{align}\label{pf_sec_2}
    I(S_{\cB}; W_{[K]}) = 0, \quad \cB \subset [N], ~ |\cB| \leq B,
\end{align}
where $S_n$ is the storage at the $n$th server, according to the given definition of Byzantine servers above.

\begin{remark}
    The main concept behind \eqref{pf_sec_2} is that if any $B$ servers can send a useful information to the user, then they can leak any message if these servers are Byzantine. This suggests that the messages should be hidden from any $B$ servers, i.e., $X$-security constraint with $X=B$.
\end{remark}

A user, who wishes to retrieve a message $\theta \in [K]$, chosen uniformly at random, sends queries $Q_n^{[\theta]} $ to the $n$th server such that none of the servers gets to know the required message index. Since the Byzantine servers share an open communication channel and the user does not know which servers are Byzantine, they need to ensure that any $B$ servers cannot decode the required message index, i.e.,
\begin{align}\label{pf_colluding}
    I(Q_{\cB}^{[\theta]};\theta)=0, \quad \cB \subset [N], ~|\cB| \leq B.
\end{align}

\begin{remark}
    If we do not allow the Byzantine servers to collude, i.e., they have an open communication channel prior to scheme initiation only, then \eqref{pf_colluding} can be dropped and the colluding requirement becomes
    \begin{align}
        I(Q_{n}^{[\theta]};\theta)=0, \quad n \in [N].
    \end{align}
\end{remark}

Upon receiving the queries, the honest servers respond truthfully with an answer, $A_n^{[\theta]}$ for server $n$, that is a function of their storage, queries, and the masking random variable, $Z_n$, i.e.,
\begin{align}
    H(A_n^{[\theta]}|S_n,Q_n^{[\theta]}, Z_n) = 0, \quad n \in \mathcal{H},
\end{align}
where $\mathcal{H} = \Bar{\cB}$ is the set of honest servers.

On the other hand, the Byzantine servers respond with answers that are arbitrary functions of their shared queries, $Q_{\cB}^{[\theta]}$, shared storage, $S_{\cB}^{[\theta]}$, the collective information of the masking random variable, $\hat{Z}_{\cB}$, and an arbitrary random variable $\gamma_{n}$, i.e.,
\begin{align}
    H(A_n^{[\theta]}|S_{\cB},Q_{\cB}^{[\theta]}, \hat{Z}_{\cB}, \gamma_{n}) = 0, \quad n \in \mathcal{B}.
\end{align}

\begin{remark}
    For comparison, in \cite{unresponsive_byzantine_1}, the Byzantine servers are only allowed to send arbitrary random variables independent of the messages and the storage and independent of other Byzantine servers, i.e., 
    \begin{align}
        I(A_n^{[\theta]}; S_n,Q_n^{[\theta]},Z_n, A_{\cB \setminus n}^{[\theta]}) = 0, \quad n \in \cB.
    \end{align}
\end{remark}

The user cannot know any information about the messages other than the required message index, i.e., 
\begin{align}
    I(W_{[K]\setminus \theta}; A_{[N]}^{[\theta]},Q_{[N]}^{[\theta]})=0.
\end{align}

\begin{remark}
    The coalition among Byzantine servers to devise a plan to harm the whole system is represented by their shared arbitrary random variables $\gamma_1, \ldots, \gamma_B$. 
\end{remark}

\section{Main Result and Consequences}

\begin{theorem}\label{main_thm}
   Given $N$ servers with $B$ of them being Byzantine. The asymptotic capacity of the symmetric PIR is given by,
   \begin{align}
       C_{BSPIR}(B,N, \infty) = 1 - \frac{4B}{N}.
   \end{align}
\end{theorem}

The following corollaries are of great importance.

\begin{corollary}
    Given the same setting as Theorem \ref{main_thm}. If there are $E$ eavesdropped links, any $T$ servers are colluding with $\max(T,E) \leq B$ and $X \leq B$ is the security parameter, then the asymptotic capacity is,
    \begin{align}
        C_{EBXTSPIR}(E,B,X,T,N, \infty) = 1 -\frac{4B}{N}.
    \end{align}
\end{corollary}

\begin{corollary}
    For a QSPIR system with $N$ databases, where $X$ of them can communicate to decode messages, $B$ are Byzantine, $E$ links are tapped by passive eavesdroppers that can listen to both uplink and downlink communications (i.e., static eavesdropper), $U$ of the databases are unresponsive, and $T$ of them are colluding to identify user-required message index, the following rate is achievable with symmetric privacy, 
    \begin{align}
        &R_Q = \nonumber \\ 
        &\begin{cases}
            2\big(\frac{N-H-M-2B-U}{N}\big),  & \parbox[t]{4.8cm} {$N-U > H+M+2B \geq \frac{N}{2}, \\ E \leq 2H+2M+2B-N,$} \\
            2\big(\frac{N-H-M-2B-U-\frac{\delta}{2}}{N}\big),  & \parbox[t]{4.8cm}{$N-U-\frac{\delta}{2}> H+M+2B \geq \frac{N}{2}, \\ E > 2H+2M+2B-N,$} \\
            \frac{N-4B-2U-E}{N},  & \parbox[t]{4.8cm} {$H+M < \frac{N}{2}, \\ 2B+U+E < H+M, \\ N > 4B+2U+E,$}\\
            \frac{N-H-2B-U-M}{N},  & \parbox[t]{4.8cm}{ $H+M < \frac{N}{2}, \\ 2B+U+E > H+M, \\ N> H+2B+U+M,$}\\
        \end{cases}
    \end{align}
     where $M = \max\{E,T\}$, $H = \max\{X,B\}$,  and $\delta = N+E -2H-2M-2B$. 
\end{corollary}

\begin{corollary}
      For a QSPIR system with $N$ databases, where $X$ of them can communicate to decode messages, $B$ are Byzantine, $|\mathcal{E}_1|\leq E$ of the uplinks and $|\mathcal{E}_2| \leq E$ of the downlinks are tapped by a passive eavesdropper, with $\mathcal{E}_1 \neq \mathcal{E}_1$ (i.e., dynamic eavesdropper). In addition, $U$ of the databases are unresponsive, and $T$ of them are colluding to identify user-required message index, the following rate is achievable with symmetric privacy,
    \begin{align}
        &R_Q = \nonumber \\
        &\begin{cases}
            2\big(\frac{N-H-M-2B-U}{N}\big) ,  & \parbox[t]{4.8cm}{ $N-U> H+M+2B \geq \frac{N}{2}, \\ E \leq 2H+2M+2B-N,$} \\
            2\big(\frac{N-H-M-2B-U-\frac{\delta}{2}}{N}\big),  & \parbox[t]{4.8cm} {$N-U-\frac{\delta}{2} > H+M+2B \geq \frac{N}{2}, \\ E > 2H+2M+2B-N,$} \\
            \frac{N-4B-2U-E}{N},  & \parbox[t]{4.8cm} {$H+M < \frac{N}{2}, \\ 2B+U+E < H+M, \\ N > 4B+2U+E,$}\\
            \frac{N-H-2B-U-M}{N},  & \parbox[t]{4.8cm} {$H+M < \frac{N}{2}, \\ 2B+U+E > H+M, \\ N > H+2B +U+2M,$}
        \end{cases}
    \end{align}
    where $M = \max\{E+B,T\}$, $H = \max\{X,B\}$, and $\delta = N+E-2H-2M-2B$.
\end{corollary}

\begin{remark}
    In comparison with the work done in \cite{Byzantine_journal,our_symmetric_quantum_byzantine_Eaves_journal,byzantine_1}, the effect of the Byzantine servers in the rate per retrieving $L$ symbols is $2B$ here instead of $3B$.
\end{remark}

\begin{lemma}
    If the Byzantine servers are allowed only to communicate before the initiation of the scheme, i.e., they do not share the queries during the scheme, then given the same setting as Theorem \ref{main_thm}, the asymptotic capacity is given by,
    \begin{align}
       C_{BSPIR}(B,N, \infty) = 1 - \frac{3B+1}{N}.
   \end{align}
\end{lemma}

\section{Achievable Scheme}\label{scheme}
Let $L = N-4B$ and the storage is given by,
\begin{align}\label{storage}
    S_n = \begin{bmatrix}
        W_{.,1} + \sum_{i=1}^B (f_1-\alpha_n)^iZ_{1,i}\\
        \vdots\\
        W_{.,L} + \sum_{i=1}^B (f_L-\alpha_n)^iZ_{L,i}
    \end{bmatrix}.
\end{align}
The queries are given by,
\begin{align}\label{queries}
    Q_n^{[\theta]} = \begin{bmatrix}
        \frac{1}{f_1 - \alpha_n}\left(e_{\theta} + \sum_{i=1}^{B} (f_1-\alpha_n)^i R_{1i}\right)\\
        \quad \vdots\\
        \frac{1}{f_L - \alpha_n} \left(e_{\theta}+ \sum_{i=1}^{B} (f_L-\alpha_n)^i R_{Li}\right)
    \end{bmatrix}.
\end{align}

First, the answers from the honest servers before providing symmetric security can be written as,
\begin{align} 
    \Bar{A}_n^{[\theta]} =& S_n^t Q_n^{[\theta]} \\ =& \sum_{\ell=1 }^L \frac{1}{f_\ell - \alpha_n} W_{\theta, \ell}  + \big(e_{\theta}^tZ_{11}+\ldots+e_{\theta}^tZ_{L,1}  \nonumber \\  & \quad +W_{.,1}^t R_{11}+ \ldots +W_{.,L}^t R_{L1}\big) \nonumber \\
    & + \sum_{i=1}^{B-1} \sum_{\ell=1}^{L} (f_{\ell}-\alpha_n)^ i \big(e_{\theta}^t Z_{\ell, i+1} + R_{\ell,1}^tZ_{\ell,i}  \nonumber \\ 
    & \quad + R_{\ell,2}^tZ_{\ell,i-1}+ \ldots+R_{\ell,i}^tZ_{\ell,1} + R_{\ell,i+1}^tW_{.,\ell} \big) \nonumber 
    \end{align}
    \begin{align}
    & + \sum_{i=0}^{B-1} \sum_{\ell=1}^{L} (f_{\ell}-\alpha_n)^ {i+B} \big(R_{\ell,i+1}^tZ_{\ell,B}  \nonumber \\ 
    & \quad+ \ldots + R_{\ell,B}^tZ_{\ell,i+1}\big) \label{answers_before_masking}. 
\end{align}

To optimize the masking random variable, we need to examine \eqref{answers_before_masking} meticulously. The first $L$ symbols are pure message symbols. The next symbol, on $\Vec{1}$ subspace, contains information about the messages, however, it is masked by the uniform random variable $e_{\theta}^t \sum_{\ell=1}^L Z_{\ell,1}$. In addition, for the next $B-1$ terms, a linear combination of new random variables $e_{\theta}^t \sum_{\ell=1}^L (f_{\ell}-\alpha_n)^i Z_{\ell,i+1}$, $i \in [1:B-1]$ is introduced to mask any side information the user can get. Finally, the last $B$ terms do not contain any newly introduced masking random variables, therefore leakage can occur. Using this observation, we need to use a masking random variable that masks the last $B$ terms. However, since any $B$ servers can be Byzantine, i.e., can communicate the masking random variable $\hat{Z}_{\cB}$ to the user and leak information, then the masking should be designed to be secure against any $B$ servers. Thus, the masking can be designed in the following way,
\begin{align}\label{masking}
    \Hat{Z}_n = \sum_{i=1}^{2B} \alpha_n^{i-1}Z'_{i}.
\end{align}
where $Z_{i}$'s are independent uniform random variables.

Thus, the transmitted answers from the honest servers can be written as,
\begin{align}
    A_n^{[\theta]} & = \Bar{A}_n^{[\theta]} + \Hat{Z}_n.
\end{align}
The complete set of answers can be written as shown in \cite{byzantine_1, Byzantine_journal, our_symmetric_quantum_byzantine_Eaves_journal} as follows,
\begin{align}\label{recieved_answers_collected}
    A^{[\theta]} =\mathtt{CSA}_{N, L,N}\begin{bmatrix}
        W_{\theta}\\
        I_{[2B]}+Z'_{[2B]}\\
        0_{2B}
    \end{bmatrix} + \Delta_B,
\end{align}
where $A^{[\theta]} = \begin{bmatrix}
        A_{1}^{[\theta]}
        \ldots
        A_{{N}}^{[\theta]}
    \end{bmatrix}^t$, and
\begin{align}
    &\mathtt{CSA}_{N, L,N} \nonumber \\ &= \begin{bmatrix}
        \frac{1}{f_1-\alpha_{1}} &\ldots&\frac{1}{f_L-\alpha_{1}}&1&\alpha_{1}&\ldots&\alpha_{1}^{N-L-1}\\
        \frac{1}{f_1-\alpha_{2}}&\ldots&\frac{1}{f_L-\alpha_{2}}&1&\alpha_{2}&\ldots&\alpha_{2}^{N-L-1}\\
        \vdots&\vdots&\vdots&\vdots&\vdots&\vdots&\vdots\\
        \frac{1}{f_1-\alpha_{{N}}}&\ldots&\frac{1}{f_L-\alpha_{{N}}}&1&\alpha_{{N}}&\ldots&\alpha_{{N}}^{N-L-1}
    \end{bmatrix},
\end{align}
and $\Delta_B$ is a vector of length $N$ with only $B$ non-zero elements that represent the Byzantine noise, $W_{\theta}$ is an $L$-length vector with the symbols of the $\theta$th message, and $I_{[2B]}$ is a $2B$-length vector with interference terms of \eqref{answers_before_masking}. 

\begin{remark}
It was shown in \cite{Byzantine_journal, our_symmetric_quantum_byzantine_Eaves_journal,byzantine_1} that a decoder exists that can decode the message symbols and the noise symbols in \eqref{recieved_answers_collected}.    
\end{remark}

\begin{remark} \label{xsecurityremark}
    If we drop the security constraint between the servers, then it is clear that $I(W_{[K]}; \Delta_B) \neq 0$. In other words, there is no way to guarantee symmetric security except by hiding the messages from any $B$ servers.
\end{remark}

\section{Illustrative Example}
Let $N = 9$ servers with $B=2$ Byzantine, thus, $L=N-4B =1$ symbol is retrieved and a possible field size is $q=11 \geq N+L =10$. We choose $\alpha_i =i$, and $f_1=10$. Let the Byzantine servers be servers $1$ and $2$. Thus, the collective answers are given by
\begin{align}
    A^{[\theta]} = \mathtt{CSA}_{9,1,9} \left( \begin{bmatrix}
        W_{\theta}\\
        I_{[2B]}\\
        0_{2B}
    \end{bmatrix} + \mathtt{CSA}_{9,1,9}^{-1} \begin{bmatrix}
        \Delta_1\\
        \Delta_2\\
        0_{7}
    \end{bmatrix} \right),
\end{align}
and by taking the inverse of the $\mathtt{CSA}_{9,1,9}$, we have,
\begin{align}
    \Hat{A}^{[\theta]} = \begin{bmatrix}
        W_{\theta} + 9 \Delta_1 + 5 \Delta_2\\
        I_1 + 5 \Delta_1 + 7 \Delta_2\\
        I_2+ 6 \Delta_1 + 3 \Delta_2\\
        I_3+ \Delta_1 + 4 \Delta_2\\
        I_4+ 10\Delta_1 + 3 \Delta_2\\
        8 \Delta_1 + 8 \Delta_2\\
        3 \Delta_1 +  \Delta_2\\
        4 \Delta_1 \\
        7 \Delta_1 + 4 \Delta_2\\
    \end{bmatrix}.
\end{align}
However, the coefficients are unknown to the user since the Byzantine servers are unknown.

\paragraph{Wrong User Estimate for Byzantine Server Locations} Let the user estimate the indices of the Byzantine servers to be $2$nd and $3$rd servers. Using the decoding structure in \cite{our_symmetric_quantum_byzantine_Eaves_journal, Byzantine_journal}, then 
\begin{align}
    \Tilde{\Phi}(\{2,3\}) &= \begin{bmatrix}
        8&10\\
        1&9
    \end{bmatrix},\\
    \Tilde{\Psi}(\{2,3\}) &= \begin{bmatrix}
        0&7\\
        4&7
    \end{bmatrix}.
\end{align}
Thus, the user computes
\begin{align}
  &\Tilde{\Phi}(\{2,3\})  \Tilde{\Psi}(\{2,3\})^{-1} \begin{bmatrix}
      4\Delta_1\\
      7\Delta_1+4\Delta_2
  \end{bmatrix} \nonumber \\
  &\quad=\begin{bmatrix}
        8&10\\
        1&9
    \end{bmatrix} \begin{bmatrix}
        8&3\\
        8&0
    \end{bmatrix}\begin{bmatrix}
      4\Delta_1\\
      7\Delta_1+4\Delta_2
  \end{bmatrix}\\
  &\quad=\begin{bmatrix}
      7\Delta_1+8\Delta_2\\
      \Delta_2
  \end{bmatrix} \neq \begin{bmatrix}
      8\Delta_1+8\Delta_2\\
      3\Delta_1+\Delta_2
  \end{bmatrix}.
\end{align}
Hence, the user knows that this choice is wrong.

\paragraph{Correct User Estimate for Byzantine Server Locations} Now if the user estimates that the Byzantine servers are $1$st, and $2$nd servers, it can be seen clearly that using the same computation as above, the last computation is equal to $\begin{bmatrix}
          8\Delta_1+8\Delta_2\\
      3\Delta_1+\Delta_2
\end{bmatrix}$. Thus, the user is able to decode the messages, since the coefficients multiplied by the noise is known now, since the Byzantine server locations are known, and they can be removed from all the message symbols and interference symbols. 

\section{Converse Proof}
Proceeding into the converse proof, we remark that we follow the same line of logic as in \cite{byzantine_tpir}. We proceed as follows,
\begin{align}
    R &= \frac{L}{\sum_{i=1}^N H(A_n^{[\theta]})} \\
    &\leq  \frac{L}{\sum_{n=1}^N H(A_n^{[\theta]}|\cq)} \label{cp_1}\\
    &= \frac{L}{(N-2B) H(A_n^{[\theta]}|\cq)} \times \frac{N-2B}{N}\label{cp_2}\\
    & \leq   \frac{L}{\sum_{n \in \mathcal{H}'} H(A_n^{[\theta]}|\cq)} \times \frac{N-2B}{N} \label{cp_3}\\
    &\leq  C_{XTSPIR}(B,B, N-2B) \times \frac{N-2B}{N} \label{cp_4}\\
    &\leq  C_{XTPIR}(B,B, N-2B) \times \frac{N-2B}{N} \label{cp_5},
\end{align}
where $\cq = \{Q_n^{[\theta]}, n \in [N], ~ \theta \in [K]\}$, \eqref{cp_2} is due to the symmetry assumption for the optimal scheme, \eqref{cp_3} follows from the fact that the answers from any $\mathcal{H}' \subset \mathcal{H}$, with $|\mathcal{H}'| = N-2B$ honest servers, the user can decode the required message, \eqref{cp_4} is due to \cite{csa} and the fact that any $B$ servers cannot know the messages or else they would be able to transmit them to the user and any $B$ servers cannot know the required message index. Finally, \eqref{cp_5} is due to the fact of dropping the symmetric privacy requirement. It was shown in \cite[Theorem~3]{csa} that by taking $\lim_{K \rightarrow \infty}$, we have,
\begin{align}
    R_{\infty} \leq \frac{N-4B}{N-2B}\times \frac{N-2B}{N}=\frac{N-4B}{N}.
\end{align}

\section{Privacy and Security Proofs}
It is clear that the queries in \eqref{queries} are secure against any $B$ servers, and the storage defined in \eqref{storage} is secure against any $B$ communicating servers. The detailed proofs for the aforementioned properties can be found in \cite{Byzantine_journal, our_symmetric_quantum_byzantine_Eaves_journal}. The only missing proof is to show that the scheme provides symmetric privacy. This can be shown using the following lemma.

\begin{lemma} 
    The scheme provided in Section~\ref{scheme} provides symmetric privacy, i.e., 
    \begin{align}\label{symmetric_privacy}
        I(W_{\theta^c}; A_{[N]}^{[\theta]}, Q_{[N]}^{[\theta]})=0.
    \end{align}
\end{lemma}

\begin{Proof}
We need to prove two different things here to show symmetric privacy. First is that the interference symbols that carry information about other messages as shown in \eqref{answers_before_masking} are secure against any Byzantine behavior after masking, and the second is to show that the Byzantine servers cannot send any useful information about the messages. To proceed in proving the first task, we need to show that the first $B$ interference terms prior to masking are independent. To show this, we need to expand the second sum in \eqref{answers_before_masking} in terms of coefficients of $\alpha_n$. To make it easier to visualize, let $\eta_{\ell,i+1} = e_{\theta}^t Z_{\ell, i+1} + R_{\ell,1}^tZ_{\ell,i} + R_{\ell,2}^tZ_{\ell,i-1}+ \ldots+ R_{\ell,i+1}^tW_{.,\ell} $. First, note that the independence of $\{\eta_{\ell,i+1}\}_{i}$ follows from the independence of $Z_{\ell,i+1}$. Then, we proceed as follows,
    \begin{align}
        &\sum_{i=0}^{B-1} \sum_{\ell=1}^{L} (f_{\ell}-\alpha_n)^ i \big(e_{\theta}^t Z_{\ell, i+1} + R_{\ell,1}^tZ_{\ell,i} + R_{\ell,2}^tZ_{L,i-1} \nonumber \\ 
        & \quad  +\ldots+ R_{\ell,i+1}^tW_{.,\ell} \big) \nonumber\\
        =&\sum_{i=0}^{B-1} \sum_{\ell=1}^{L} \sum_{k=0}^{i} {i\choose k} f_{\ell}^{i-k} \alpha_n^k \eta_{\ell,i+1}\\
        =& \sum_{i=0}^{B-1} \sum_{k=0}^{i} {i\choose k} \alpha_n^k \underbrace{\sum_{\ell=1}^{L}   f_{\ell}^{i-k}  \eta_{\ell,i+1}}_{\mu_{k,i}}\\
        =&   \big({0 \choose 0} \mu_{0,0}+ {1 \choose 0} \mu_{0,1} + \ldots {B-1 \choose 0} \mu_{0,B-1}\big) \nonumber \\ 
        &  + 
        \alpha_n \big({1 \choose 1} \mu_{1,1} + \ldots + {B-1 \choose 1} \mu_{1,B-1}\big) \nonumber \\
         &+  \alpha_n^2 \big( {2 \choose 2} \mu_{2,2} + \ldots {B-1 \choose 2}\mu_{2,B-1}\big) \nonumber\\ &+ \ldots + \alpha_n^{B-1} \big({B-1 \choose B-1} \mu_{B-1,B-1}\big)\\
         =& \big( \sum_{\ell=1}^L \eta_{\ell,1} + f_{\ell}\eta_{\ell,2}+ \ldots + f_{\ell}^{B-1} \eta_{\ell,B}\big) + \alpha_n \big(\sum_{\ell=1}^L {1 \choose 1}\eta_{\ell,2} \nonumber \\ 
         &+ {2 \choose 1}f_{\ell}\eta_{\ell,3}+\ldots + {B-1 \choose 1}f_{\ell}^{B-2} \eta_{\ell,B}\big) \nonumber \\
         &+ \alpha_n^2 \big(\sum_{\ell=1}^L {2 \choose 2}\eta_{\ell,3}+ {3 \choose 2}f_{\ell}\eta_{\ell,4}+\ldots \nonumber \\ &+ {B-1 \choose 2}f_{\ell}^{B-3} \eta_{\ell,B}\big) + \ldots + \alpha_n^{B-1} \big(\sum_{\ell=1}^L \eta_{\ell,B}\big)\label{independence_equation}.
    \end{align}
    Now, it is important to note that in the $\Vec{1}$ subspace we have $\eta_{\ell,1}$, that does not exist in any other subspaces of $\Vec{\alpha}, \ldots, \Vec{\alpha}^{B-1}$, similarly $\eta_{\ell,2}$ does not exist in the subspaces $\Vec{\alpha}^2, \ldots, \Vec{\alpha}^{B-1}$, which illustrates the independence of interference terms in $\Vec{\alpha}^{k}$ subspaces where $k \in \{0,\ldots,B-1\}$. Moreover, by the one-time pad theorem, due to the independence of $Z_{\ell,i+1}$, they are independent from anything beside $Z_{\ell,i+1}$. Now note that
    \begin{align}   I(&W_{\theta^c};A^{[\theta]}_{[N]},Q^{[\theta]}_{[N]})=I(W_{\theta^c};Q^{[\theta]}_{[N]})+I(W_{\theta^c};A^{[\theta]}_{[N]}|Q^{[\theta]}_{[N]}) \\
    &=I(W_{\theta^c};A^{[\theta]}_{[N]}) =I(W_{\theta^c};A^{[\theta]}_{\mathcal{H}},A^{[\theta]}_{\cB}) \\
    &=I(W_{\theta^c};W_{\theta},I_1+Z'_1,I_2+Z'_2,\Delta_{\cB}) \label{info_answers}
    \end{align}
    where $I_1+Z'_1$ is the first $B$ terms of $I_{[2B]}+Z'_{[2B]}$ and $I_2+Z'_2$ are the remaining terms, and last equality follows from \eqref{recieved_answers_collected} and the correctness of the scheme. Now, consider the case of the Byzantine servers sending queries they received, back to the user. Then, using the independence of queries as before, it can be seen \eqref{info_answers} gives,
    
    \begin{align} I(W_{\theta^c};A^{[\theta]}_{[N]},Q^{[\theta]}_{[N]})=I(W_{\theta^c};W_{\theta},I_1+Z'_1,I_2+Z'_2)=0,
    \end{align}
    
    by the symmetry of the scheme without Byzantine servers. If the Byzantine servers are to send their storage to the user, similarly from the independence of storage random variables $Z_{\ell,i}$, \eqref{info_answers} gives,
    \begin{align} &\!\!I(W_{\theta^c};A^{[\theta]}_{[N]},Q^{[\theta]}_{[N]})=I(W_{\theta^c};W_{\theta},I_1+Z'_1,I_2+Z'_2,S_{\cB}) \!\!\\ &=I(W_{\theta^c};W_{\theta},I_1+Z'_1,I_2+Z'_2)+I(W_{\theta^c};S_{\cB}) =0,
    \end{align}
    which is why $X$-security was implemented, as pointed in Remark \ref{xsecurityremark}. If the Byzantine servers are to send $\gamma_{\cB}$ to the user, then again by the independence of $\gamma_{\cB}$, \eqref{info_answers} gives,
    \begin{align}  &\!\!I(W_{\theta^c};A^{[\theta]}_{[N]},Q^{[\theta]}_{[N]})=I(W_{\theta^c};W_{\theta},I_1+Z'_1,I_2+Z'_2,\gamma_{\cB}) \!\!\\ &=I(W_{\theta^c};W_{\theta},I_1+Z'_1,I_2+Z'_2)+I(W_{\theta^c};\gamma_{\cB}) =0.
    \end{align}

    Now, if the Byzantine servers send their share of symmetric masking variable $\hat{Z}$, that is $\hat{Z}_{\cB}$, then, 
    \begin{align} &\!\!\!I(W_{\theta^c};A^{[\theta]}_{[N]},Q^{[\theta]}_{[N]})=I(W_{\theta^c};W_{\theta},I_1+Z'_1,I_2+Z'_2,\hat{Z}_{\cB}) \!\!\\  &=I(W_{\theta^c};W_{\theta},I_1+Z'_1,\hat{Z}_{\cB})\nonumber\\
    &\quad+I(W_{\theta^c};I_2+Z'_2|W_{\theta},I_1+Z'_1,\hat{Z}_{\cB}) \\
    &=I(W_{\theta^c};I_2+Z'_2|W_{\theta},I_1+Z'_1,\hat{Z}_{\cB}) \label{independence_proof} \\
    &=H(W_{\theta^c}|W_{\theta},I_1+Z'_1,\hat{Z}_{\cB})\nonumber\\ 
    &\quad - H(W_{\theta^c}|I_2+Z'_2,W_{\theta},I_1+Z'_1,\hat{Z}_{\cB}) \\
    &=H(W_{\theta^c})\nonumber+H(\hat{Z}_{\cB}|W_{\theta},I_1+Z'_1,I_2+Z'_2)\\
    &\quad -H(W_{\theta^c},\hat{Z}_{\cB}|W_{\theta},I_1+Z'_1,I_2+Z'_2) \label{independence_proof_2}\\
    &=H(W_{\theta^c})\nonumber+H(\hat{Z}_{\cB})-H(W_{\theta^c})\\
    &\quad -H(\hat{Z}_{\cB}|W_{[K]},I_1+Z'_1,I_2+Z'_2) \label{independence_proof_3} \\
    &=I(\hat{Z}_{\cB};W_{[K]},I_1+Z'_1,I_2+Z'_2) \\ &=I(\hat{Z}_{\cB};W_{[K]})+I(\hat{Z}_{\cB};I_2+Z'_2)\nonumber\\
    &\quad+I(\hat{Z}_{\cB};I_1+Z'_1|I_2+Z'_2)\label{independence_proof_4} \\
    &=0,
    \end{align}
    where \eqref{independence_proof} follows from the independence of messages from $Z'_{[2B]}$ and from each other, \eqref{independence_proof_2} follows from the same reasoning, and the use of chain equalities, \eqref{independence_proof_3} follows from $I(\Hat{Z}_{\cB};I_1+Z'_1,I_2+Z'_2) = 0$ shown by independence of $\eta_{\ell,i+1}$ in \eqref{independence_equation}, and \eqref{independence_proof_4} follows from the same logic. Then, using data processing inequality, since $\hat{Z}_{\cB}, S_{\cB}, \gamma_{\cB}, Q_{\cB}$ are independent from each other and since $A^{[\theta]}_{\cB}$ is a function of these four variables; we see that $I(W_{\theta^c};A^{[\theta]}_{[N]},Q^{[\theta]}_{[N]})=I(W_{\theta^c};W_{\theta},I_1+Z'_1,I_2+Z'_2,\Delta_{\cB}) 
    \leq I(W_{\theta^c};W_{\theta},I_1+Z'_1,I_2+Z'_2,\hat{Z}_{\cB}, S_{\cB}, \gamma_{\cB}, Q_{\cB})
    =0.$
\end{Proof}
\newpage

\bibliographystyle{unsrt}
\bibliography{references.bib}

\end{document}